# Unilateral Antidotes to DNS Poisoning

Amir Herzberg and Haya Shulman

Department of Computer Science
Bar Ilan University
Ramat Gan, Israel
{amir.herzberg,haya.shulman}@gmail.com

**Abstract.** We investigate defenses against DNS cache poisoning focusing on mechanisms that can be readily deployed unilaterally by the resolving organisation, preferably in a single gateway or a proxy. DNS poisoning is (still) a major threat to Internet security; determined spoofing attackers are often able to circumvent currently deployed antidotes such as port randomisation. The adoption of DNSSEC, which would foil DNS poisoning, remains a long-term challenge.

We discuss limitations of the prominent resolver-only defenses, mainly port and IP randomisation, 0x20 encoding and birthday protection. We then present two new (unilateral) defenses: the *sandwich antidote* and the *NAT antidote*. The defenses are simple, effective and efficient, and can be implemented in a gateway connecting the resolver to the Internet. The *sandwich antidote* is composed of two phases: poisoning-attack *detection* and then *prevention*. The *NAT antidote* adds entropy to DNS requests by switching the resolver's IP address to a random address (belonging to the same autonomous system). Finally, we show how to implement the birthday protection mechanism in the gateway, thus allowing to restrict the number of DNS requests with the same query to 1 even when the resolver does not support this.

## 1 Introduction

Correct and efficient operation of the DNS is essential for the operation of the Internet. However, there is a long history of vulnerabilities and exploits related to DNS; for some of the early works, see the seminal papers of Vixie [30] and Bellovin [7, 8].

In the recent years, the most significant attack on the Domain Name System is *DNS poisoning* by a spoofing attacker. The spoofer tries to provide the DNS resolver with misleading mappings (e.g., map VIC-Bank.com to an IP address controlled by the attacker), by sending a fake (spoofed) response to a domain name query. The DNS poisoning scenario by a spoofing adversary typically assumes either an open recursive resolver, i.e., one that provides services to clients outside of its network, or a compromised client on the local network (LAN), e.g., a zombie; the DNS poisoning model is in Figure 1. DNS poisoning can be used as a building block facilitating many other attacks, such as the injection of malware, phishing, website hijacking/defacing and denial of service.

When the DNS resolver receives a DNS response, it usually follows the recommendations in [18] and checks that the validation fields match the fields in one of the pending DNS requests. The DNS request contains several validation fields, e.g., transaction ID, which are also copied to the DNS response by the authoritative name server. The local DNS resolver that issued the DNS query validates that those fields appear correctly in the DNS response. If all the fields in the DNS response are correct, the response is cached and then sent to the client that issued the request. Otherwise, if one of the fields is incorrect, the DNS resolver ignores response. Once the DNS response is cached, the attacker has to wait until the TTL (time to live) expires so that it can initiate the attack again.

Although poisoning attacks on DNS were known to be devastating, this threat was believed to be impractical, since frequently accessed domain names typically reside in the cache of the DNS resolver, thus preventing the attacker from poisoning those domains of interest. Furthermore, if the legitimate response from the authentic DNS server arrives before the forgery sent by the attacker, forgery attempt fails, as the resolver will cache the first response and ignore the rest.

This situation changed when Kaminsky presented an improved attack [20, 10], with two critical improvements. The first improvement was to control the time at which the resolver sends queries (to which the attacker wishes to respond), by sending to the resolver queries for a (non-existing) host name, e.g., with a random or sequential prefix of the domain name. The second improvement was to add, in the spoofed responses sent to the resolver, a type NS DNS record (specifying a new name for the domain name server) and/or a type A 'glue' DNS record (specifying the IP address of the new domain name server). These records poison the resolver's entries for a specific host in the victim's domain, e.g., the victim's name server itself. Hence, if the attack succeeds once (for one record), the adversary controls the entire name space of the victim. If the attack fails for a given host name (prefix), the attacker can repeat with new (random) prefix.

Using these two improvements attackers can often poison the DNS entry for the victim domain (e.g., VIC-Bank.com) within few seconds, when the only unpredictable field in the DNS response is the 16-bit ID field, thus allowing devastating attacks on many Internet applications (see [20, 10]).

As a result of Kaminsky attack, it became obvious that changes are needed to prevent DNS poisoning. Indeed, major DNS resolvers were quickly patched to support *source port randomisation*, i.e., use and validate random source ports for each request or at least for each destination IP. Resolvers were also improved to support *birthday protection*: prevent or limit[1] duplicate concurrent requests. Both of these defenses were proposed by Bernstein already in [9].

However, as noted by [21, 14], a determined attacker with sufficient bandwidth, e.g., controlling a large amount of compromised machines on the Internet, could still send a sufficient number of responses to have a forged DNS response accepted with high probability. Furthermore, port randomisation is often annihilated due to port-mangling by NAT devices between the resolver and the Internet [17, 26, 12].

---

[1] Complete prevention of duplicate queries may have significant overhead on popular resolver implementations, hence most implementations only limit duplicates.

Additional, easy to deploy defense requiring changes to the local DNS resolver only, RFC 5452 [18], specifies that DNS resolvers should, where possible, not only choose a random source port, but also choose a random source IP address and a random authoritative server IP address for each query. The resolver should then validate that the same IP addresses are used in the response. If the resolver uses a set of $N_S$ IP addresses, and the authoritative name server uses a set of $N_D$ IP addresses, then the space of identifiers is increased by a factor of $N_S \cdot N_D$. However, the impact is usually modest, as $N_S$ is often only one, and $N_D$ is at most three.

Indeed, recent studies, [19, 16], indicate that DNS is not well protected against poisoning attacks and that the vast majority of organisations with an Internet presence are still vulnerable to DNS poisoning attacks. In particular, the ongoing attacks on DNS infrastructure, e.g., AT&T, Comcast and Rollingstone [32, 11, 23], motivate inspection of the 'easy-to-deploy', unilateral defenses such as those against spoofing adversaries.

Cryptographic defenses, e.g., DNSSEC [4, 6, 5], offer protection against a stronger man-in-the-middle adversary, and are thus a preferable alternative over defenses against spoofers. However, their deployment remains to be seen due to the significant changes that they introduce to the current DNS infrastructure. Recent survey results, [19, 16], reveal that some fundamental capabilities required for adoption of DNSSEC, e.g., support of queries over TCP and support of EDNS0 [29], are not fully deployed. Furthermore, common to the defenses against MitM adversaries is the requirement for a cooperation and support of the mechanism by both parties to the DNS transaction, which is a significant overhaul. In contrast, unilateral defenses against spoofers allow an organisation or an ISP to integrate the defense, without relying on the support by the other end to the DNS transaction.

In this work, we focus on antidotes to DNS poisoning that require modification to the local DNS resolver only, which, preferably, can also be implemented in a router/firewall machine connecting the resolver to the Internet (we discuss advantages of firewall based defenses in Section 1.1).

Our proposed defenses meet all the proposed design guidelines of [24]: (1) the prevention techniques should require no change to the DNS protocol; (2) should not introduce service disruption; (3) the solution should be completely backward compatible with existing DNS servers, and transparent to users; and finally (4) it should make poisoning attacks infeasible.

Our first technique is the *sandwich antidote*, presented in Section 3. This is an efficient and simple procedure, based on a two stage defense: upon the receipt of a forged DNS response, an attack is detected; then, the attack is prevented by discarding the 'malicious' DNS responses and accepting only a valid authentic DNS response.

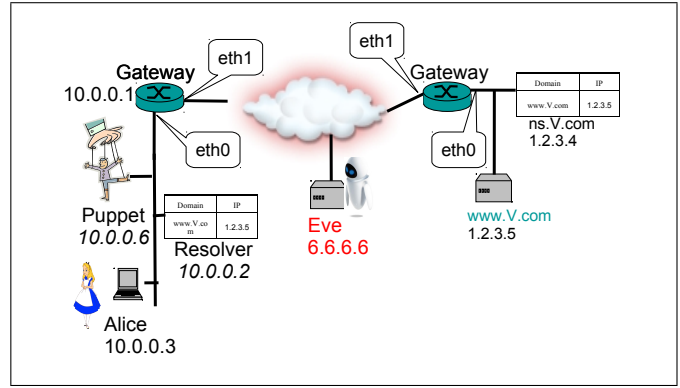

**Fig. 1.** Simple configuration for DNS poisoning by a spoofing adversary (Eve) on the Internet: client Alice uses a resolver connected to the Internet. The resolvers make queries to authoritative name servers, via the Internet. We consider a blind/spoofing adversary Eve, connected to the Internet. The adversary may also control some 'puppet' connected to the same local area network as Alice and the resolver.

Our second technique is the *NAT antidote*, Section 4. The NAT antidote extends the existing Network Address Translation devices and adds entropy to DNS requests, thus significantly increasing the amount of forged packets that the attacker is required to generate in order to produce the correct *forged* DNS response.

We also show how to implement the birthday protection in firewall. The mechanism restricts the number of outgoing DNS requests for the same query to 1, even if the local DNS resolver does not limit concurrent requests for the same resource record.

We implemented our proposed defenses in linux gatway, and tested their compatibility by querying the real DNS servers on the Internet. Implementing solutions in the gateway has several advantages, over implementations in the resolver itself, which we discuss in Section 3.

### 1.1 Firewall-Based Defense Mechanisms

Anti-poisoning defenses implemented in firewall have several advantages over defenses in the resolver:

- *Integration Challenges:* resolver software may not be amenable to modification due to complexity or due to it being proprietary, while modifications in the firewall are simpler, as firewalls already have built in tools to capture packets. The modification can be made in a small user-space program (as we did for our experimental validation of the prevention techniques), which is much simpler than modifying the resolver.

– *One Firewall Protects Many:* one firewall, e.g., of ISP, can protect all the DNS resolvers, e.g., of ISP's clients, without the need to integrate the changes in all the local DNS resolvers.
– *Modular Design:* If a DNS server is replaced, or a new server is added, there is no need to integrate the modification each time.
– *Security Feature:* for firewall vendors, adding another security feature is an important added value.

### 1.2 Contributions and Organisation

We present two practical and efficient defense mechanisms against DNS poisoning, that require changes only to the local DNS resolver and protect the local DNS resolver against poisoning attacks by spoofing adversaries. Our defenses can be implemented in the firewall. The *sandwich antidote* (Section 3) is simple to implement and integrate in a gateway which connects the network to the Internet, requires modest resources, and can provide sufficient entropy to make poisoning infeasible. The *NAT antidote* (Section 4) is very simple to deploy, with almost negligible overhead, and in many cases, provided a significant number of client IP addresses are available, can significantly improve defense against poisoning.

In Section 2 we discuss and compare recently proposed antidotes to DNS poisoning.

Finally, in Section 5, we show how birthday protection can be efficiently implemented in a firewall, which is significant since most existing resolvers only *limit* the amount of duplicates, e.g., to 200.

We present the implementations of our proposed defenses in Linux based firewall.

## 2 Proposed Antidotes to DNS Poisoning

In this section we briefly review proposed anti-poisoning defenses against spoofers, that require integration on the side of the local DNS resolver only. These techniques can be broadly categorised as follows: (1) mechanisms that increase entropy in DNS packets (subsection 2.1), and (2) mechanisms that inspect the DNS responses to detect forgery (subsection 2.2).

### 2.1 Entropy Increasing Mechanisms

Unilateral entropy increasing mechanisms, most notably: source port randomisation (SPR) [9], source/destination IP address randomisation (IPR) and DNS 0x20 encoding [14], add more randomness to DNS packets, in order to make it more difficult for the spoofing adversary to craft a valid DNS response, that would get accepted *and* cached by the local DNS resolver. In order to produce a successful forgery the attacker has to guess *all* the values in the validation fields correctly. The more random values pertain in the DNS requests, the lower the probability of the attacker to produce a successful guess. Entropy increasing mechanisms do not try to identify forgery attacks and respond to them, but attempt to increase the difficulty of producing a successful forgery.

Unfortunately, the number of fields in the DNS packet, that could be used to add randomness, is limited: the transaction ID field in the DNS packet, the source port, source/destination IP addresses, and the choice of upper/lower case for the query (since DNS is case-insensitive; this is the field used by DNS 0x20 encoding).

**Port Randomisation** Following to Kaminsky attack, [20], the need to patch the DNS resolvers to send DNS requests from random ports became apparent. Using random source ports adds another 16 bits of entropy, resulting in a search space of $2^{16} \times 2^{16} = 2^{34}$ bits, which makes successful poisoning significantly more difficult to achieve.

Unfortunately, source port randomisation (SPR) is resource intensive, and as noted in [13, 15] may be inappropriate for busy DNS servers or embedded devices. In addition, as pointed out by [14], determined attackers can overcome source port randomisation, by sending large amounts of traffic, e.g., from many zombie computers, thus covering all the search space and eventually producing a valid DNS response. Therefore, enhancing DNS security using SPR and transaction ID alone may not suffice.

Furthermore, many (or most) DNS resolvers are located behind NAT devices, and as [12, 10] noted, NAT devices that use sequential assignment of external ports, may expose (even the patched) local DNS resolvers, to poisoning attacks, since NAT could reduce source port randomisation implemented by the local resolvers.

**IP Randomisation** IP addresses are known to be a scarce resource on the Internet. IP address is composed of 32 bits, resulting in at most $2^{32}$ possible addresses. Due to this shortcoming, the DNS resolvers are often allocated a single IP address, and authority DNS servers are allocated at most 3 IP addresses. This can increase the search space of the attacker by at most a factor of $1 \times 3$, which does not offer sufficient protection. In addition, the DNS resolvers that are located behind NAT devices, which is the typical case, lose the IP randomisation, even if they are allocated several IP addresses.

**DNS 0x20 Encoding** Dagon *et al.* [14] present an innovative technique, 0x20-encoding, for improving DNS defense by increasing entropy of DNS queries against poisoning by spoofed responses. The technique is based on an observation that domain names are case insensitive, however, most authoritative servers copy the string of the domain name from the incoming request to the response they send back, *exactly* as sent - preserving the case of each letter. They suggest to randomly toggle the case of letters of which the domain name consists, and validate them in response. If the domain name $d$ contains $l(d)$ alphabetic characters, this increases the space of identifiers by factor of $X(d) = 2^{l(d)}$, e.g., $X(\textsc{www.google.com}) = 2^{12}$ and $X(\textsc{a9.com}) = 2^4$.

Note that in Kaminsky-style attacks, the query is for a non-existing domain name chosen by the attacker, e.g., to poison addresses in the domain GOOGLE.COM, an attacker may issue a query for $r$.GOOGLE.COM where $r \in_R \{0, \ldots, 9\}^8$ is a random string of 8 digits, resulting in a domain name with only 9 letters, i.e., factor of only $X = 2^9$; namely, 0x20 encoding is less effective for domain names containing few letters - which are often the most important domains, e.g., the Top Level Domains (TLDs) such as .COM and .UK.

Although 0x20 encoding, as presented by Dagon *et al.*, introduces significant extra entropy to DNS requests, an attacker may still be able to poison 'high value' domain names with a rather small factor of poisoned responses. Therefore, there is need in alternative or additional technique to protect the TLDs and other domain names containing only few letters, e.g., A9.COM.

## 2.2 Forgery Detection Mechanisms

The 'forgery detection' mechanisms follow two approaches: the collaborative approach and techniques from machine learning. According to the collaborative approach the authenticity of a DNS response is validated by distributing the DNS requests across hosts in the system, e.g., [22, 25, 28], or by consulting a set of trusted peers, [31], and then, e.g., taking the majority answer. CoDNS, ConfiDNS and DoX, [22, 25, 31], send the requests to several peers in a peer-to-peer network and accept the first DNS response; if the first response is forged, it is still accepted. DepenDNS, [28], queries several DNS resolvers, and accepts the DNS response of the majority. DepenDNS relies on open recursive DNS resolvers to obtain DNS responses; open recursion DNS services are known to expose DNS to attacks. Furthermore, recently, [2] showed that DepenDNS does not protect DNS against poisoning attacks. The common shortcomings of the collaborative approach are most notably the performance penalty, i.e., additional processing and communication delays, that they introduce to every DNS request, even when the system is *not* under attack, and the significant infrastructure that is required for deployment. In addition, techniques that are based on distributing the DNS request to several nodes, e.g., [22], are also exposed to cache poisoning attacks, as the first DNS response that arrives is accepted.

A recent technique, by Antonakakis *et al* [3], employs mechanisms from machine learning to identify suspicious IP addresses. Specifically, [3] designed a centralised poisoning detection system called Anax, which is based on the observation that DNS records direct users to a known set of NS records, while poisoned records redirect users to new IP addresses, outside of the victim's address space. However, deployment requires trust in one central entity that should be consulted to establish authenticity of the DNS responses. In addition, this mechanism also introduces delays and may have false positives, e.g., if an authority DNS server was moved to a new IP address for load distribution.

# 3 The Sandwich Antidote to DNS Poisoning

Both categories, the increasing entropy (Section 2.1) and the forgery detection (Section 2.2) mechanisms, have different shortcomings and most importantly: the DNS cache poisoning problem is not yet solved, thus motivating further investigation of anti-poisoning defenses.

In this section we present an anti-poisoning defense technique, the sandwich antidote, designated to run in a gateway (or a proxy), behind which the local DNS resolver is located, and should filter DNS traffic. The sandwich antidote is applied only to DNS packets and is based on first detecting and then activating the prevention module, to counter poisoning attempts. As a result, the mechanism does not impose performance overhead, and is only applied when DNS cache poisoning attack is detected. The sandwich antidote maintains a table that stores all outbound DNS requests, prior to forwarding them. It also keeps track of the inbound DNS responses, and matches them against the pending DNS requests. If a corresponding DNS request exists, and the DNS response is correct, i.e., all the validation fields match the corresponding values in the DNS request, then the entry is removed from the table and the response is forwarded to the DNS resolver.

A poisoning attack is detected when one of the validation fields, e.g., transaction ID, in a DNS response does not match[2] the corresponding value in the pending DNS request, in which case, the prevention module is activated. During the course of the poisoning attack many invalid DNS responses for some DNS request may arrive. Among these incorrect responses a valid response may appear, however, with high probability this can be a forged response generated by the attacker. Therefore, once an attack was detected, i.e., an incorrect DNS response was received for some pending query, the mechanism should *not* rely on that (seemingly correct) DNS response, since it can be merely a successful forgery sent by the attacker.

Once activated, the sandwich antidote issues three DNS requests, substituting the original DNS request sent by the resolver: (1) a DNS request for the requested resource record, prepended with some random string, e.g., if the original DNS request was for *x.y.com*, then after receiving an invalid DNS response for query *x.y.com*, the mechanism should issue a request for *randomString1.x.y.com*; (2) a DNS request for the original resource record, i.e., *x.y.com* as above; (3) a DNS request for *randomString2.x.y.com*, where *randomString2* is a randomly selected string, and *x.y.com* is the RR as appeared in the original DNS request.

The sandwich antidote expects to receive correct DNS responses to all three requests and in the *same order* in which the requests were sent. Specifically, it checks that the DNS responses to the above requests are correct, and arrive in the same order, in which the requests were sent. This mechanism is based on the observation that it may be possible to generate a single correct DNS response, to a potentially adversarial query. However, generating three correct DNS responses, where the first and third are random, should not be feasible, let alone ensuring that all three are received in the same order in which they were sent. The authentic responses to those three DNS requests above should be: (1) an nxdomain (i.e., with high probability the hostname $randomString1.x.y.com$ does not exist) or an NS RR, i.e., a referral to a name server lower in hierarchy, e.g., from $ns.com$ to $ns.y.com$; (2) an $A$ RR (IP address for $x.y.com$, or an NS RR, i.e., a referral to the DNS lower in hierarchy; (3) same as (1) above, i.e., an nxdomain or an NS RR. Note that as a result of this 'order preserving' mechanism, where the original query is between two queries prepended with a random prefix, we coin this mechanism the 'sandwich antidote'.

---

[2] Note that when generating a DNS response, the DNS servers copy the validation fields from the DNS request accurately, thus it is guaranteed that the validation fields should appear correct in the authentic DNS response.

Once correct DNS responses arrive in the required order, for all three DNS requests, the mechanism removes the pending DNS requests from the table and returns the DNS response to the DNS resolver. The sandwich mechanism ignores the responses if they arrive in an incorrect order, or if the validation fields do not match. The diagram describing the functionality of the sandwich antidote is in Figure 3. Due to length restrictions, the pseudo-code appears in the full version of the paper.

Note that similarly to other unilateral defenses, such as SPR or 0x20 encoding, the sandwich antidote does not offer protection when it is implemented in a resolver (or a firewall) which uses a higher level resolver as a forwarder, e.g., receives services from the resolver of an ISP.

In subsequent sections we present a detailed design of the sandwich mechanism. We discuss the additional overhead that our mechanism inflicts on the gateway and analyse the impact of the sandwich mechanism on the probability of the attacker to produce a successful forgery. We also show that our mechanism does not expose the DNS resolver (or its clients) to denial of service (DoS) attacks.

### 3.1 Detailed Design

In order to keep track of the outbound DNS requests and to record the poisoning attempts, the mechanism maintains two tables: the table $T$ that stores all outbound DNS requests (for which a valid DNS response has not arrived yet), and the table $A$ that maintains the DNS requests for which poisoning attempts were detected.

Table $T$ is illustrated in Table 1. The table which is indexed by a hash function $h$ applied on the query field of the DNS request. Upon arrival of a DNS request Req, the hash $h(Req.query)$ is stored in table $T$ and is mapped to the validation fields, i.e., source port $srcPort$, source/destination IP addresses $srcIP$ and $dstIP$, and DNS 0x20 encoding bits $X$:
$T[h(Req.query)]=(srcPort||srcIP||dstIP||X)$. Namely, the table $T$ is composed of the indices column containing a digest of the query, and a column for each 'validation' field, and a serial number (which is used to locate the entry in $T$ from table $A$). Table $T$ can contain entries with the same index[3], i.e., when several DNS requests for the same RR were issued simultaneously.

Table $A$ is illustrated in Table 2. Table $A$ maintains DNS requests which were issued once forgery attempt was detected for some DNS request. Table $A$ is composed of indices column h(Req.query), a column containing a fingerprint, i.e., the result of a PRF (pseudo-random function) $f$ applied on the entropy fields T[h(Req.query)]=$f_K(srcPort\|srcIP\|dstIP\|X)$, and a column containing a serial number pointing to the corresponding entry in $T$ (required to generate the DNS response once responses for corresponding entries in $A$ were received). Specifically, once an invalid DNS response arrives for an existing, pending DNS request, the mechanism issues three DNS requests, and stores them in A.

The diagram, in Figure 3, describes the functionality of the sandwich antidote. Upon receipt of a DNS response, in step (1), the mechanism checks if an entry for that query exists in table $A$, in 2, step (2).

If no matching entry exists in $A$, however a corresponding DNS requests is stored in $T$, Table 1, step (3), i.e., the DNS request was issued, the DNS response is checked (the validation fields are compared against those in the pending DNS request), step (4). If the response is correct, it is sent to resolver, and the pending DNS request is removed from table $T$. If the response is incorrect, e.g., the destination port in the response does not match the source port in the DNS request, attack is detected, and prevention module is triggered. Specifically, the mechanism issues three DNS requests, such that the DNS request containing the original query, is sent between two other DNS requests that contain random strings prepended to the original query: (a) \$1.\|Resp.query, (b) Resp.query, (c) \$2.\|Resp.query, s.t., request (b) contains the original query as appeared in the DNS request, and (a) and (c) contain the original request prepended with distinct random strings \$1 and \$2. These queries are stored in table $A$, and further DNS responses for that query will be processed against these entries in $A$.

If a matching entry exists in $A$, an attack was detected, i.e., at least one DNS response containing wrong values in the validation fields was received; the mechanism should not accept the DNS response as is, but should apply a special processing described next. In this case, the mechanism first checks that the validation fields match against those in the corresponding entry in $A$, step (5), and that the DNS response arrived in the correct order, i.e., according to the order in which the DNS requests were sent, step (6).

We implemented steps (5) and (6) as a state machine, Figure 2. The state machine is activated following to attack detection and transmission of three DNS requests (as above). The state machine transits to subsequent states following to successful receipt of the DNS response, i.e., in order and correct. The state machine halts when step 4., Figure 2, is reached. Then the DNS response, for the corresponding DNS request stored in $T$, is sent to the DNS resolver, and the respective entries are removed from $A$ and $T$.

---

[3] Following to Kaminsky attack, DNS resolvers were patched to support birthday protection, we extend more on this in Section 5, and we also show a gateway based mechanism to restrict the number of outstanding DNS requests to 1 to prevent the birthday paradox.

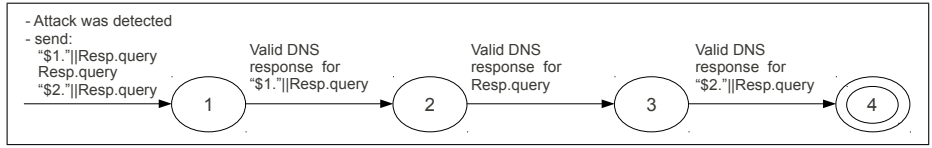

**Fig. 2.** The state machine of the sandwich antidote, that corresponds to steps (5) and (6) in Figure 3.

| Index | Serial number | Transaction ID | Source IP | Destination IP | Source port | 0x20 encoding |
|---|---|---|---|---|---|---|
| h('WWw.GOoGLe.COM') | $i$ | 12567 | 1.2.3.4 | 5.6.7.8 | 55555 | 110110110111 |
| h('WwW.yAhOO.coM') | $i+1$ | 35783 | 1.2.3.4 | 7.8.9.7 | 61345 | 10101011001 |
| h('Www.msn.COm') | $i+2$ | 22344 | 1.2.3.3 | 2.6.8.3 | 24580 | 100000110 |

**Table 1.** Sample entries, containing the DNS request sent by the DNS resolver, in the table $T$ maintained by the sandwich antidote mechanism running at the gateway; for simplicity the entries are presented using ASCII characters.

| Index | Serial number | Fingerprint |
|---|---|---|
| h('wAKjfruEHa.WWw.GOoGLe.COM') | $i.1$ | $f_K(12567||1.2.3.4||5.6.7.8||44563||011000011111101101101111)$ |
| h('wWw.gOOGlE.cOm') | $i.2$ | $f_K(23455||1.2.3.4||5.6.7.8||1089||010011101010)$ |
| h('OknfDEJFNa.wwW.gOogle.CoM') | $i.3$ | $f_K(12577||1.2.3.4||5.6.7.8||54333||1000111110001010000101)$ |

**Table 2.** Sample entries in table $A$, containing the DNS request sent by the sandwich antidote mechanism, running at the gateway, once forgery attempt was detected for DNS query 'www.google.com'; for simplicity the entries are presented using ASCII characters.

### 3.2 Sandwich Implementation in Firewall

We implemented the sandwich antidote in C and ran it as a user-space program on Linux Netfilter (kernel 2.6) operating system. We added two rules to iptables firewall, one to capture all outbound packets destined to port 53 and another to capture all inbound packets originating from port 53. The kernel passed the captured packets to the implementation, which queued them, and applied to each packet the processing described in Section 3.1.

We tested that the sandwich antidote does not impose overhead to DNS or other traffic and that it does not 'break' the DNS functionality in two settings: without poisoning attack and when under attack.

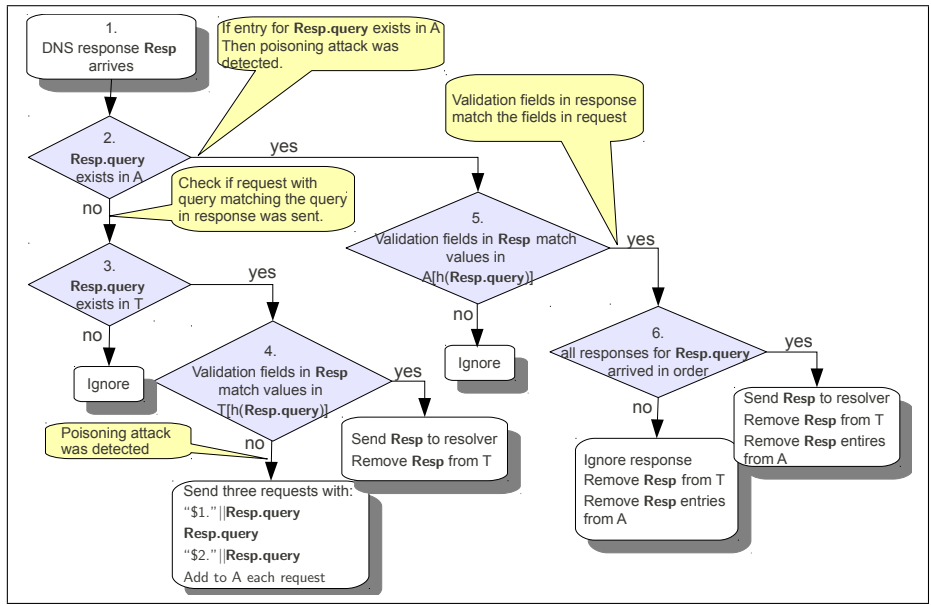

**Fig. 3.** A flow of the defense mechanism in firewall of the local resolver. The flow describes the steps taken by the firewall upon arrival of each DNS response.

**Compatibility with DNS Infrastructure** When our resolver was configured to use a forwarder, in some cases the responses did not arrive in correct order. This is due to the fact that many domains were already in cache of the forwarder, while the two random queries did not exist. Thus the two responses for random queries would arrive after the response to the original DNS request. To adapt the mechanism to such settings, where the local resolver uses a forwarder, we modified the implementation to issue the requests twice. Specifically, if the responses do not arrive in correct order, the same three DNS requests (random, original, random) are issued again.

**Efficiency** When no attack is detected, the mechanism does not impose delays on other (non-DNS) traffic. The delays on DNS packets were in order of few microseconds. To reduce even these (negligible) delays we added a slight optimisation to our original design: we modified the firewall rule to duplicate the outbound DNS packets, then to forward, in parallel, the DNS packet to the destination (on the Internet) and a copy thereof for processing to the user-space implementation.

## 4  The NAT Antidote

RFC 5452 [18] recommends that when sending a query, DNS resolvers should use a random source IP address (from a list of IP addresses allocated to the resolver) and a random destination IP address (from the list of addresses of authoritative servers for the domain). Upon receiving a response, resolvers should validate it matches the query, in their source and destination IP addresses (as well as in the resolver's port, DNS query identifier, and case-sensitive query). Validation of IP addresses increases the entropy and makes poisoning by spoofed responses harder. Specifically, if $n_R$ ($n_A$) is the number of addresses used for resolver (respectively, authority DNS), then the improvement is by factor $n_R \times n_A$.

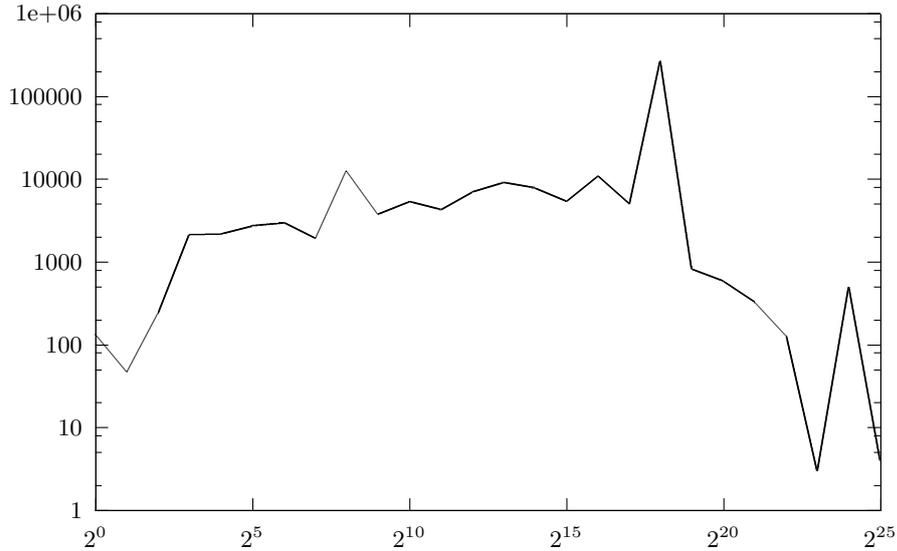

**Fig. 4.** Statistics of the top 100,000 domains according to Alexa, and the size of addresses blocks allocated to them. The X-asis denote the size of the IP addresses block and Y-axis denote the number of networks that have that number of IP addresses.

However, typically, only a small number of IP addresses $n_R, n_A$ are used for resolver and authority DNS; e.g., Dagon et al. [14] mention typical values of $n_R = 1$ and $n_A = 3$. Clearly, increasing $n_R$ and $n_A$ would improve resistance to poisoning; however, IP addresses is a scarce, expensive resource, hence allocating additional IP addresses to the DNS servers is hard to justify. On the other hand, often, a domain may have multiple IP addresses used for other purposes (not resolver), e.g., an ISP may have many IP addresses allocated to clients, and a company may have special IP addresses allocated to different publicly-available servers. The NAT antidotes takes advantage of such addresses. Specifically, we show how to add IP addresses to DNS resolver without allocating additional IP addresses. The idea is to reuse IP addresses already allocated to the network. Hence this method works for networks that have a large set of public IP addresses; fortunately, this holds for networks of many organisations and ISPs; see

Figure 4 summarising the IP range blocks for top 100, 000 domains according to Alexa [1], which we used in order to form a dataset. We gathered this information by running a script on the list of domains (freely available on Alexa); the script employed the whois service in order to obtain the IP address block of the domain. According to our survey the median is at $2^{11}$ which indicates that most domains have a sufficiently large block of IP addresses. The deployment of NAT antidote only requires modifications in the router(s) connecting the organisation to the Internet, much like commonly used network address translation (NAT) and firewall devices. Indeed, sometimes all that is required is to use existing overloading many-to-many NAT functionality.

If necessary, implementation is similar to NAT, using tools such as iptables. Figure 5 illustrates typical operation of the NAT antidote.

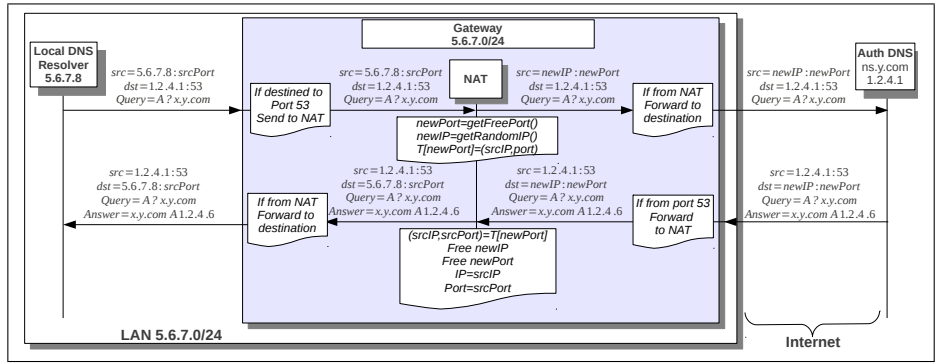

**Fig. 5.** The NAT Antidote, implementing the defense against DNS cache poisoning by increasing the amount of source IP addresses, used by the local DNS resolver, to the number of addresses allocated to the network. The NAT antidotes connects a network (e.g., 5.6.7.0/24) to the Internet. The NAT maps a DNS request, sent by the resolver to authority DNS, to a random IP address in the network; the NAT also changes the source port to some random, free port number, and saves this entry to a table. This allows the NAT to change the destination IP and port, in the DNS response, to the values that were originally used by the local DNS resolver.

### 4.1 NAT Implementation in Firewall

The NAT antidote is a variant of many-to-many, overloading NAT functionality, i.e., mapping a set of internal IP addresses to a set of external IP address. Upon receiving a DNS request (to port 53), the firewall would perform a special type of Network Address Translation function. Specifically, it changes the IP address, of the outbound DNS requests, to a *random* IP address in the available range, stores the mapping (between the source IP, source Port, destination IP), and forwards the packet to the DNS server. When the DNS response arrives, the firewall locates the corresponding entry according to the destination IP and port, and changes the destination port and IP address to those that were used by the resolver. It then removes the stored entry and sends the DNS response; see Figure 5.

## 5 Birthday Paradox and Protection

Originally, DNS servers did not restrict the number of simultaneous, multiple DNS requests for the same IP address. This allowed a mathematical paradox, known as the "Birthday Paradox", to reduce the number of DNS responses required for a successful poisoning. To exploit the birthday paradox one has to send a sufficient number of queries to a DNS server, while sending an equal number of forged DNS responses at the same time. The greater the number of outgoing DNS requests, the greater the probability that the attacker will match one of those requests with a forged DNS response. As an example assume that only the transaction ID (16 bits long) is randomised in the DNS request, and that the attacker sends $n$ responses to a DNS request; the success probability of a match of one the responses with the DNS request is $\frac{n}{2^{16}}$. If the attacker issues $n$ requests for the same domain, and sends $n$ responses, then the probability of a collision of one of the responses with one of the DNS request (with a random transaction ID) is: $1 - \left(1 - \frac{1}{2^{16}}\right)^{n(n-1)/2}$. For instance, for $n = 300$ there is a 50% success probability for a match and for $n = 700$ there is almost 100% success, while without the birthday paradox the success probability is $700/2^{16} = 0.01$, e.g., see [27].

Following to Kaminsky attack, many DNS resolvers were also improved to support *birthday protection*: prevent or limit [4] duplicate concurrent requests. In subsequent section we suggest a mechanism that completely prevents the birthday paradox.

### 5.1 Birthday Protection

Spoofing attacker can significantly increase its success probability of poisoning the cache of local DNS resolver by issuing a number of DNS requests for the same resource record, thus taking advantage of the birthday paradox. Birthday protection limits the number of concurrent DNS requests for the same record. However, not all DNS servers implement birthday protection, or sufficiently restrict the number of DNS queries. We suggest to implement the birthday protection mechanism in the firewall, without requiring modification to the local resolver itself. The idea is to limit multiple duplicate DNS requests, for the same RR (resource record) to one, by having the firewall return a single DNS response, i.e., the first one that arrives, and to ignore the rest.

The birthday protection mechanism should run at the gateway (or a proxy DNS server, through which the local resolver will issue DNS requests and receive

---

[4] Complete prevention of duplicate queries may have a significant overhead, hence many implementations only limit duplicates, e.g., to 200.

DNS responses). The gateway should capture DNS requests and responses, by adding appropriate rules to the firewall, and should keep track of the outbound DNS requests. Specifically, when a packet destined to port 53 enters the network interface card (NIC) $eth_0$, (as in Figure 1), the firewall captures and queues it for processing by a waiting userspace birthday protection mechanism implementation. The mechanism maintains a table (see Table 3 for sample entries) of DNS requests and when a new request arrives it is stored in the table. The mechanism then checks if a DNS request with the same query already exists in table. If the request is new and does not exist in a table, the *first* flag is set to 1, i.e., this is the first DNS request. Then the birthday protection mechanism forwards it to the designated recipient; otherwise, the *first* is set to 0 and the request is discarded. When a DNS response arrives from 53 the firewall passes it to the birthday protection mechanism. If a matching entry exists with a *flag* set to 1 the mechanism processes the response (otherwise ignores it). Taking only the response that matches the first query is important to prevent the attack exploiting the birthday paradox. For each entry in a table with a query that matches the query in the DNS response, the mechanism constructs a DNS response with the fields, e.g., port, source/destination IP, that correspond to the values in table, and uses the same *answer* field value from the DNS response that it received; it then sends the responses and removes the corresponding entries from the table. All future DNS responses for that RR will be ignored (since no matching entry exists in the table).

It may seem that the mechanism should only return the response to one of the matching DNS requests, and ignore the rest of the requests. However, depending on the implementation of the DNS resolver, when several clients make requests for the same resource record (RR), the resolver does not return the first (cached) response to the other clients, but waits for the corresponding response to arrive. Therefore, our mechanism crafts a matching DNS response to every pending DNS request. Therefore, our mechanism returns a matching DNS response to each pending entry.

| Index | Query ID | source IP | destination IP | source Port | 0x20 Encoding | first |
|---|---|---|---|---|---|---|
| h('www.google.com') | 12567 | 1.2.3.4 | 5.6.7.8 | 55555 | 1101101101 | 1 |
| h('www.google.com') | 2234 | 1.2.3.4 | 5.6.7.9 | 3112 | 1101100011 | 0 |

**Table 3.** Sample entries in the table maintained by the birthday protection mechanism; for simplicity the entries are presented using ASCII characters.

## 6   Conclusions

Currently, many DNS resolvers are still vulnerable to DNS poisoning attacks by determined adversaries, only requiring from them the ability to spoof packets. DNS poisoning can be used for a wide range of devastating attacks, hence, it is essential to develop interim solutions, to ensure security until the long-term cryptographic DNS-security mechanisms are widely deployed. Preferably, such interim antidotes to DNS poisoning should require changes only in the resolver, or, better yet, only in the gateway connecting the resolver to the Internet.

We investigated unilateral defenses against the DNS cache poisoning by spoofing adversaries, and presented new and improved mechanisms. Our central contribution is the *sandwich antidote* to DNS poisoning, which operates in two phases: detecting and then preventing poisoning attacks. We also presented the *NAT antidote*, which enhances DNS security by increasing the entropy in DNS packets for most subnets by a factor of $2^{11}$, by picking a random source IP address from a pool of addresses available to the organisation. These solutions can be easily deployed in gateways (we present proof of concept code), to provide immediate defense against DNS poisoning.